\documentclass[prd,superscriptaddress,amsfonts,amssymb,amsmath,showpacs,twocolumn,showkeys]{revtex4-1}

\usepackage{amsmath}
\usepackage{epsfig}
\usepackage{palatino}
\usepackage{graphics}
\usepackage[colorinlistoftodos]{todonotes}
\usepackage{color}
\usepackage{graphicx}
\usepackage[font={footnotesize,it}]{caption}
\usepackage[colorlinks=true,
            linkcolor=black,
            urlcolor=blue,
            citecolor=blue]{hyperref}

\def\be{\begin{equation}}
\def\ee{\end{equation}}
\def\bea{\begin{eqnarray}}
\def\eea{\end{eqnarray}}

\begin{document}


\title{ Quasinormal Modes and Greybody Factors of $f(R)$ gravity  minimally coupled to a cloud of strings in $2+1$ Dimensions}

\author{Ali \"{O}vg\"{u}n}
\email{ali.ovgun@pucv.cl}
\affiliation{Instituto de F\'{\i}sica, Pontificia Universidad Cat\'olica de
Valpara\'{\i}so, Casilla 4950, Valpara\'{\i}so, CHILE.}

\affiliation{Physics Department, Arts and Sciences Faculty, Eastern Mediterranean University, Famagusta, North Cyprus via Mersin 10, TURKEY.}
\affiliation{Physics Department, California State University Fresno, Fresno, CA 93740,
USA.}
\affiliation{Stanford Institute for Theoretical Physics, Stanford University, Stanford,
CA 94305-4060, USA}

\author{Kimet Jusufi}
\email{kimet.jusufi@unite.edu.mk}
\affiliation{Physics Department, State University of Tetovo, Ilinden Street nn, 1200,
Tetovo, MACEDONIA.}
\affiliation{Institute of Physics, Faculty of Natural Sciences and Mathematics, Ss. Cyril and Methodius University, Arhimedova 3, 1000 Skopje, MACEDONIA.}

\date{\today }

\begin{abstract}
In this paper we have studied the propagation of a massless scalar field minimally coupled  to a  black hole with the source of cloud of strings in $2+1$  $f(R)$gravity theory.  In particular we have found analytical results for the decay rate, reflection coefficient, greybody factors, as well as the black hole temperature. On the other hand, our quasinormal modes analyses reveals stability in the propagation of a massless scalar field in the cloud of strings black hole spacetime. However, under a suitable choose of parameters an instability is found. Furthermore based on the Bekenstein conjecture on the quantization of the surface area of the black hole we find the minimal surface area associated to the black hole horizon which is in agreement with Bekenstein's proposal.
\end{abstract}


\maketitle

\section{Introduction}
From a theoretical point of view, black holes are one of the most fascinating objects predicted by Einstein's theory of relativity.  There are many interesting phenomena related to the black hole physics, in particular the Hawking radiation process \cite{hawking1,hawking2}. In 1970s, Bekenstein  had an amazing insight, namely he conjectured that the black-hole entropy should be proportional to the area of the event horizon \cite{Bekenstein1,Bekenstein2}. Then in 1974 Hawking made a major step forward toward understanding the quantum nature of black holes by arguing that black holes should radiate quanta of particles due to the quantum effects in the curved spacetime \cite{hawking1,hawking2}. 

In a seminal paper, Parikh and Wilczek discovered that Hawking radiation can be described in terms of the tunneling formalism due to the conservation of energy. Moreover, they showed that the exact spectrum is not precisely thermal \cite{wil1,par1}. Then Corda showed that it is not hardly continuous so that it provides the link between quasinormal modes (QNMs) and Hawking radiation (HR) \cite{corda1,corda2,corda3}. Furthermore, another important contribution was made by Hod who argued that the quantization of a black hole area can be estimated by the asymptotic value of its QNMs  \cite{Hod1,Hod2,Hod3,Hod4,Hod5,Hod6,Hod7,Hod8,Hod9,Hod10}. In 2008, Maggiore corrected Hod's conjecture by calculating the area of the horizon of a Schwarzschild black hole as $\Delta A=8\pi l_{pl}^2$, where $l_{pl}$ is the Planck length in contrast with the Hod's result of $\Delta A=4\ln(3) l_{pl}^2$ \cite{Mag}. Since then, black holes have been widely studied in fact, physicist tend to believe that black holes might provide the key to understand the ultimate quantum theory of gravity \cite{Kubiznak1,Kubiznak2,ovgun0,ovgun1,ovgun2,ovgun3,ovgun4,ovgun5,mann1,mann2,mann3,mann4,ovgun6,sin1,sin2,sin3,sin4,sin5,sin6,sin7,bir1,bir2,bir3,bir4,bir5,bir6}. In this direction one can point out the information loss paradox which continues to attract a lot of interest.

On the  other hand, greybody factors and QNMs have attracted a lot of interest in the recent years \cite{qnm0,qnm1,qnm2,qnm3,qnm4,qnm5,qnm6,qnm7,qnm8,qnm9,qnm10,qnm11,qnm12,qnm13,qnm14,qnm15,qnm16,qnm17,qnm18,qnm19,qnm20,qnm21,qnm22,qnm23,qnm24,qnm25,qnm26,qnm27,qnm28,qnm29,qnm30,qnm31,qnm32,qnm33,qnm34,qnm35,qnm36,qnm37,qnm38,qnm39,qnm40,qnm41,qnm42,qnm43,qnm44,qnm45,qnm46,qnm47,qnm48,qnm49,qnm50,qnm51,qnm52,qnm53,qnm54,qnm55,qnm56,qnm57,qnm58,qnm59,qnm60,qnm61,qnm62,qnm63,qnm64,qnm65,qnm66,qnm67,qnm68,qnm69,qnm70,qnm71,qnm72,qnm73,qnm74,qnm75,qnm76,qnm77,qnm78,qnm79,qnm80,qnm81,qnm82,qnm83,qnm84,qnm85,qnm86,qnm87,qnm88,qnm89,qnm90,qnm91,qnm92,qnm93,qnm94,qnm95,qnm96,qnm97,qnm98,Schiappa1,Schiappa2,liu1,liu2,liu3}. Due to the nontrivial geometry of the spacetime metric Hawking radiation deviates from a pure thermal nature. Some part of the waves will be reflected back to the black hole due to effective potential barrier as a result of the gravitational field. One can use greybody factors to calculate the reflection coefficient, absorption cross section and the flux of these fields. On the other hand, QNMs are used to test the stability of astrophysical objects or fields in a given spacetime background geometry. Furthermore QNMs play a crucial role in the study of gravitational waves which has been recently discovered by LIGO \cite{LIGO}. Quasinormal modes are considered as fingerprints which help us to directly identify the black hole existence by detecting gravitational waves. It is amazing that we can estimate the black hole mass, or angular momentum parameter, by detecting gravitational waves. 

In this paper we study the greybody factors and the QNMs by performing a massless scalar field perturbations of $f(R)$ gravity  minimally coupled to a cloud of strings in $2+1$ dimensions (CSBHs). Since the paper of Banados, Teitelboim and Zanelli (BTZ) black hole \cite{Banados:1992wn}, black hole solutions in $2+1$ dimensions have became more important according to many reasons.  One of the reason to study on the lower dimensions is to simplify calculations.  Without adding new sources, it is not possible to find the black hole solutions in spacetime lower than four dimensions. For this reason, authors added source of cloud of strings  and minimally coupled to $f(R)$ gravity \cite{Mazharimousavi:2015sfo}. The energy momentum of the cloud of strings in $2+1$ dimensions are at the order of $\frac{1}{r}$, on the other hand the energy momentum of the electromagnetic sources are of the order $\frac{1}{r^2}$. The cloud of strings make weaker the singularity at $r=0$, with compared to the other sources, that motive the authors to study on it. 

The paper is organized as follows. In Section II, we review the $2+1$  dimensional CSBHs. In Section III, we shall consider a massless scalar field perturbation. In Section IV, we find an analytical solution to the radial wave equations. Among other thing we shall discuss the stability and calculate the greybody factors such as the reflection coefficient, absorption cross section and the flux.

\section{$2+1$ dimensional CSBHs}
\label{3dBH}
The action of a $f(R)$ gravity coupled to the cloud of strings in $2+1$ dimensional spacetime is given by \cite{Mazharimousavi:2015sfo}
\begin{equation}
S=\frac{1}{16 \pi G}\int \sqrt{-g}\,f(R) d^3x+S_{GN},
\end{equation}
where the second term is the Nambu-Goto action
\begin{equation}
S_{GN}=\int_{\Sigma} m \sqrt{-h} \,d\lambda^0 d\lambda^1.
\end{equation}

Note that $m$ is a positive constant related to the tension of the string and $\left(\lambda^0, \lambda^1 \right)$ are the string parameters. The energy-momentum tensor is written as
\begin{equation}
T^{\mu \nu}=\rho \frac{\Sigma^{\mu \alpha}\Sigma^{\nu}_{\alpha}}{\sqrt{-h}},
\end{equation}
where $\rho$ is the energy density, $\Sigma^{\mu\nu}$ is the spacetime bi-vector.  Furthermore $h=\det(h_{ab})$ gives the determinant of the induced metric
\begin{equation}
h_{ab}=g_{\mu \nu} \frac{\partial x^\mu}{\partial \lambda^a}\frac{\partial x^\nu}{\partial \lambda^b}.
\end{equation}

Letting $f(R)=R$, it was found a static, circularly symmetric spacetime that was also found in the context of the massive gravity theory given as follows
\cite{Mazharimousavi:2015sfo,ricardo}
\be\label{metric}
ds^2=-f(r)dt^2+f(r)^{-1}dr^2+r^2d\theta^2,
\ee
where the metric function $f(r)$ is given by
\begin{equation}
f(r)=-M+2\xi r,
\end{equation}%
and the Ricci scalar is  \begin{equation}
R=-\frac{4\xi }{r}.
\end{equation}
Note that the $\xi $ is a constant which can be positive (black hole) or negative (naked singularity). The event
horizon for the positive $\xi $ which represents black hole is located at%
\begin{equation}
r_{h}=\frac{M}{2\xi }.
\end{equation}

Furthermore, we can write the metric function in terms of the horizon as follows:

\begin{equation}
f(r)=2\xi( r-r_{h}).
\end{equation}

In the next section we shall investigate a scalar perturbation in the CSBHs spacetime metric given by Eq. (\ref{metric}).
\bigskip 

\section{Scalar Perturbation of $2+1$ dimensional CSBHs}

In this section, we solve a scalar field equation, namely Klein-Gordon equation on the background of $2+1$ dimensional CSBHs:
\begin{equation} \label{KG}
\frac{1}{\sqrt{-g}} \partial_{\mu} ( \sqrt{-g} g^{\mu \nu} \partial_{\nu} \Phi ) =0,
\end{equation}
using the anzatz,
\begin{equation}
\Phi =  e^{- i \omega t} e^{i m \theta} R(r).
\end{equation}
The radial equation component of Eq. \eqref{KG} is


\begin{equation}
R'' + \left( \frac{1}{r} + \frac{f'}{f} \right) R' + \left( \frac{\omega^2}{f^2} - \frac{m^2}{r^2 f} \right) R = 0,
\end{equation}
where the prime $'=\frac{d}{dr}$.

After we define $R(r)=\frac{\zeta(r)}{\sqrt{r}}$, and the tortoise coordinate $r_{*}$ coordinate $r_*$ given as
\begin{equation}
r_* = \int \frac{dr}{f(r)}=\frac{\ln(r-r_h)}{2 \xi},
\end{equation}
it is possible to find a  Schrodinger-like equation given by
\begin{equation}
\left(\frac{d^2 }{dr_*^2} + \omega^2 - V_{eff}(r) \right) \zeta(r_*) = 0.
\end{equation}

Note that the effective potential $V_{eff}(r)$ is given by,
\begin{equation}
V_{eff}(r) = \frac{ m^2 f(r)}{r^2} -\frac{f(r)^2}{4 r^2}+f'\frac{f(r)}{2r}.
\end{equation}

Furthermore from Eq. (13) one can see that
\begin{equation}
r-r_h=\exp\left(2\, \xi \,r_*\right),
\end{equation}
for the case of $r \rightarrow r_h$, $r_* \rightarrow - \infty$, and for the case of $r \rightarrow \infty$, $ r_* \rightarrow \infty $. We make coordinate transformation from $r$ to $z$ to analytically obtain the solution of the wave equation:
\begin{equation}
 z = \left(\frac{r- r_h}{ r} \right),
\end{equation}
then Eq.(12) becomes,
\begin{equation}
z(1-z) \frac{d^2 R}{dz^2} + (1-z) \frac{d R}{dz} + P(z) R =0,
\end{equation}
Here,
\begin{equation}
P(z) = \frac{A}{z} + \frac{B}{-1+z} - C
\end{equation}
where,
\begin{equation}
A=\frac{\omega^2}{4 \xi^2}  ; \hspace{1.0cm} B = -\frac{\omega^2}{4\xi^2}; \hspace{1.0cm} C =\frac{m^2}{2r_h \xi}.
\end{equation}

It is noted that the new horizon is located at $z=0$ and there is an infinity at $z =1$.

\section{Greybody and QNMs of $2+1$ dimensional CSBHs}

In this section, we study the greybody factors and QNMs of $2+1$ dimensional CSBHs.

\subsection{Exact solution in terms of hypergeometric functions}

To calculate the greybody, we first define $R$ as,
\begin{equation}
R(z) = z^{\alpha} (1-z)^{\beta} F(z),
\end{equation}
the radial equation given in Eq. (18) reduces to following equation
\begin{equation}\notag
z(1-z) \frac{d^2 F}{dz^2} + \left(1 + 2 \alpha - (1+ 2 \alpha + 2\beta )z \right) \frac{d F}{dz} 
\end{equation}
\begin{equation}
+\left(\frac{\bar{A}}{z} + \frac{\bar{B}}{-1+z} - \bar{C}     \right) F =0,
\end{equation}
where
\begin{eqnarray}\notag
\bar{A} &=& A + \alpha^2,\\\notag
\bar{B} &=& B + \beta - \beta^2,\\\notag
\bar{C} &=& C+(\alpha+ \beta)^2.
\end{eqnarray}

Demanding that $\bar{A} =\bar{B}=0$, we can determine the $\alpha$ and $\beta$ as follows
\begin{eqnarray}
\alpha_{\pm} &=& \pm \frac{  i \omega}{2\,\xi },\\
\beta_{\pm} &=& \frac{1}{2} \pm {\frac {i\sqrt {{\omega}^{2}-{\xi}^{2}}}{2\, \xi}}.
\end{eqnarray}

If we introduce the following parameters $c, a, b$ as follows
\begin{eqnarray}
c &=& 1+ 2 \alpha,\\
a &=&  \alpha + \beta+i \sqrt{C},\\
b &=&  \alpha +  \beta-i \sqrt{C},
\end{eqnarray}
then Eq. (18) resembles the hypergeometric differential equation  which is of the form \cite{Horowitz:1993jc},
\begin{equation}
z(1-z) \frac{d^2 F}{dz^2} + (c  - (1+a + b )z) \frac{d F}{dz} -ab  F =0.
\end{equation}

Using a simple arithmetic we find
the following relations for these parameters 
\begin{eqnarray}
c_{1,2} &=& 1 \pm \frac{i \omega}{\xi} ,\\
a_{1,2} &=& \frac{1}{2} \pm i\left( \frac{\omega+\sqrt{\omega^2-\xi^2}}{2 \xi} +\frac{m}{\sqrt{M}} \right) ,\\
b_{1,2} &=& \frac{1}{2}\pm i\left( \frac{\omega+\sqrt{\omega^2-\xi^2}}{2 \xi} +\frac{m}{\sqrt{M}} \right).
\end{eqnarray}
Note that $\omega^2 \geq \xi^2$ for large frequencies.
The general solution for the radial part on the other hand  is given in terms of the hypergeometric function $F(z)$ given by,
\begin{equation}
F(z) = C_1F(a,b;c;z) +C_2 z^{1-c}F(a-c+1,b-c+1;2-c;z)
\end{equation}
where $C_1$ and $C_2$ are constants.

\subsection{Near Horizon Solutions}

Now, we study the near horizon solutions of the wave equation. Let us recall Eq. (17) by rewriting it in the following form
\begin{equation}
z = 1 - \frac{r_h}{r}
\end{equation}
and 
\begin{equation}
r_* = \frac{1}{2 \xi}\ln\left(\frac{2 \xi r-M}{2\xi}\right)
\end{equation}
in the ``tortoise'' coordinate.
If one approach to horizon $r \rightarrow r_h$, it follows $z \rightarrow 0$ and the solution becomes the function $
R(z) = A_1 e^{\alpha ln z}+ A_2 e^{-\alpha ln z}.
$
Choosing $\alpha=\alpha_{-}$ and $\beta=\beta_{+}$, it reduces to
\begin{equation}
R(z  \rightarrow 0)= A_1 z^{\frac{-i\omega}{2\xi}}  + A_2 z^{\frac{i\omega}{2\xi}},
\end{equation}
with $z$ written as follows
\begin{equation}
z \approx  \frac{ r - r_h}{ r_h}=\frac{\exp\left(2 \xi r_*  \right)}{r_h}.
\end{equation}
The hypergeometric function has two linearly independent solutions with the generic radial solution for $R(z)$ which can be written as,
\begin{equation}\notag
R(z) = A_1 z^{\frac{-i\omega}{2\xi}} (1-z)^{\frac{1}{2}+{\frac {i\sqrt {{\omega}^{2}-{\xi}^{2}}}{2\, \xi}}}F\left(a,b;c;z\right) 
\end{equation}
\begin{equation}
 +A_2 z^{\frac{i\omega}{2\xi}}(1-z)^{\frac{1}{2}+{\frac {i\sqrt {{\omega}^{2}-{\xi}^{2}}}{2\, \xi}}} F\left(a-c+1,b-c+1;2-c;z\right),
\end{equation}

where $A_1$ and $A_2$ are constants and the ingoing wave is shown by first term and the outgoing wave is defined for second term \cite{Chandrasekhar:1975zza}. Then we impose the condition that there should be only purely ingoing wave at the horizon so that we choose $A_1 \neq 0$ and $A_2=0$. Therefore the solution becomes \cite{Chandrasekhar:1975zza,zib},
\begin{eqnarray} \label{sol}
R(z \rightarrow 0) =  A_1 z^{\frac{-i\omega}{2\xi}} (1-z)^{\frac{1}{2}+{\frac {i\sqrt {{\omega}^{2}-{\xi}^{2}}}{2\, \xi}}} \\ \times F\left(a,b;c;z\right) \notag.
\end{eqnarray}

\subsection{ Solutions at asymptotic region}
To find the far field solution $r \gg r_H$ we notice that $f(r) \sim 2\xi r$ and the mass term dominates over the frequency and angular momentum, and thus the differential equation for the radial part and after it is expanded, the Euler's equation is formed as follows:
\begin{equation}
r^2 R'' + 2 r R'+kR =0,
\end{equation}
where $k=-B$.
Then we calculate the solution of the Euler's equation
\begin{equation} \label{40}
R(r) = D_1 \left( \frac{r_h}{r} \right)^{\rho_1}+D_2 \left( \frac{r_h}{r} \right)^{\rho_2},
\end{equation}
with 
\begin{equation}
\rho_1=   \frac{ 1 + \sqrt{ 1 + 4 B} }{2} =  \beta,
\end{equation}
and
\begin{equation}
\rho_2=   \frac{ 1 - \sqrt{ 1 + 4 B} }{2} =  (1-\beta).
\end{equation}

In terms of the tortoise coordinate $r_*$ the solution takes the form
of plane waves when the root $r_h$ is complex.

\subsection{Matching the solutions}

In this section, we match the solution of near horizon in Eq. (\ref{sol}) and the solution of asymptotic region (the large $r$ limit or the $z \rightarrow 1$) in Eq. (\ref{40}). To this end, we use the transformation of the hypergeometric function in Eq. (\ref{40}) of the form
$$
F(a,b,c,z) = \frac{ \Gamma(c) \Gamma(c-a-b)}{\Gamma(c-a) \Gamma(c-b)} F(a,b;a+b-c+1;1-z) 
$$
\begin{equation}
 +(1-z)^{c-a-b}\frac{ \Gamma(c) \Gamma(a+b-c)}{\Gamma(a) \Gamma(b)} F(c-a,c-b;c-a-b+1;1-z).
\end{equation}

Thus we find the general solution of asymptotic region in terms of the parameters $a,b,c$ as follows:
$$
R(z) = A_1 z^{\frac{-i\omega}{2\xi}} (1-z)^{\frac{1}{2}+{\frac {i\sqrt {{\omega}^{2}-{\xi}^{2}}}{2\, \xi}}} \frac{ \Gamma(c) \Gamma(c-a-b)}{\Gamma(c-a) \Gamma(c-b)} F_1 $$
\begin{equation}
+ A_1  z^{\frac{-i\omega}{2\xi}} (1-z)^{\frac{1}{2}-{\frac {i\sqrt {{\omega}^{2}-{\xi}^{2}}}{2\, \xi}}} \frac{ \Gamma(c) \Gamma(a+b-c)}{\Gamma(a) \Gamma(b)} F_2,
\end{equation}
where we have introduced
\begin{eqnarray}\notag
F_1&=&F(a,b;a+b-c+1;1-z),\\ \notag
F_2 &=&F(c-a,c-b;c-a-b+1;1-z).
\end{eqnarray}

After we take the limit $ z \rightarrow 1$,  it's easy to see that the solution for $R(z)$ at asymptotic region reads:
\begin{equation}\notag
R(z \rightarrow 1) =  A_1  (1-z)^{\frac{1}{2}+{\frac {i\sqrt {{\omega}^{2}-{\xi}^{2}}}{2\, \xi}}} \frac{ \Gamma(c) \Gamma(c-a-b)}{\Gamma(c-a) \Gamma(c-b)} 
\end{equation}
\begin{equation}
+ A_1 (1-z)^{\frac{1}{2}-{\frac {i\sqrt {{\omega}^{2}-{\xi}^{2}}}{2\, \xi}}} \frac{ \Gamma(c) \Gamma(a+b-c)}{\Gamma(a) \Gamma(b)}.
\end{equation}

By introducing the coefficients $D_1$ and $D_2$:
\begin{equation}
D_1 = A_1 \,\frac{ \Gamma(c) \Gamma(c-a-b)}{\Gamma(c-a) \Gamma(c-b)}, 
\end{equation}
\begin{equation}
D_2 = A_1 \, \frac{ \Gamma(c) \Gamma(a+b-c)}{\Gamma(a) \Gamma(b)},
\end{equation} 
and make use of the fact that
\begin{equation}
1 - z = \frac{r_h}{r},
\end{equation}
the radial solution $R(r)$ for $ r \rightarrow \infty$ (or $ z \rightarrow 1$) is written as follows:
\begin{equation}\notag
R(r \rightarrow \infty ) = D_1  r_h^{\frac{1}{2}+{\frac {i\sqrt {{\omega}^{2}-{\xi}^{2}}}{2\, \xi}}} \exp\left(-i \omega r_* \sqrt{1-\frac{\xi^2}{\omega^2}}-\xi r_*\right)
\end{equation}
\begin{equation}
+ D_2 r_h^{\frac{1}{2}-{\frac {i\sqrt {{\omega}^{2}-{\xi}^{2}}}{2\, \xi}}} \exp\left(i \omega r_* \sqrt{1-\frac{\xi^2}{\omega^2}}-\xi r_*\right).
\end{equation}

The first (second) term represents the ingoing (outgoing) wave, respectively. The presence of $r_h$ terms is not important they are absorbed into $D_1$ and $D_2$.

\subsection{Reflection Coefficient}

The reflection coefficient $\Re$ can be defined as,
\begin{equation}
\Re =  \left|\frac{D_1}{D_2} \right|^2.
\end{equation}

Note that we have used the identities for $\Gamma$ functions given by,
\begin{eqnarray}
|\Gamma( i y )|^2 &=&  \frac{ \pi } { y \sinh ( \pi y ) },\\
|\Gamma( \frac{1}{2} + i y )|^2 &=& \frac{ \pi} { \cosh ( \pi y ) },\\
|\Gamma( 1 + i y )|^2 &=& \frac{ \pi y } { \sinh ( \pi y )}.
\end{eqnarray}
We find 
\begin{equation} \label{rr}
\Re =  \frac{\cosh[\pi (\frac{\omega-\sqrt{\omega^2-\xi^2}}{2\xi}+\frac{m}{\sqrt{M}})]\cosh[\pi (\frac{\omega-\sqrt{\omega^2-\xi^2}}{2\xi}-\frac{m}{\sqrt{M}})]}{\cosh[\pi (\frac{\omega+\sqrt{\omega^2-\xi^2}}{2\xi}+\frac{m}{\sqrt{M}})]\cosh[\pi (\frac{\omega+\sqrt{\omega^2-\xi^2}}{2\xi}-\frac{m}{\sqrt{M}})]}.
\end{equation}
 and then plot the reflection coefficient $\Re$ in Fig.1.
\begin{figure}[h!]
\center
\includegraphics[width=0.45\textwidth]{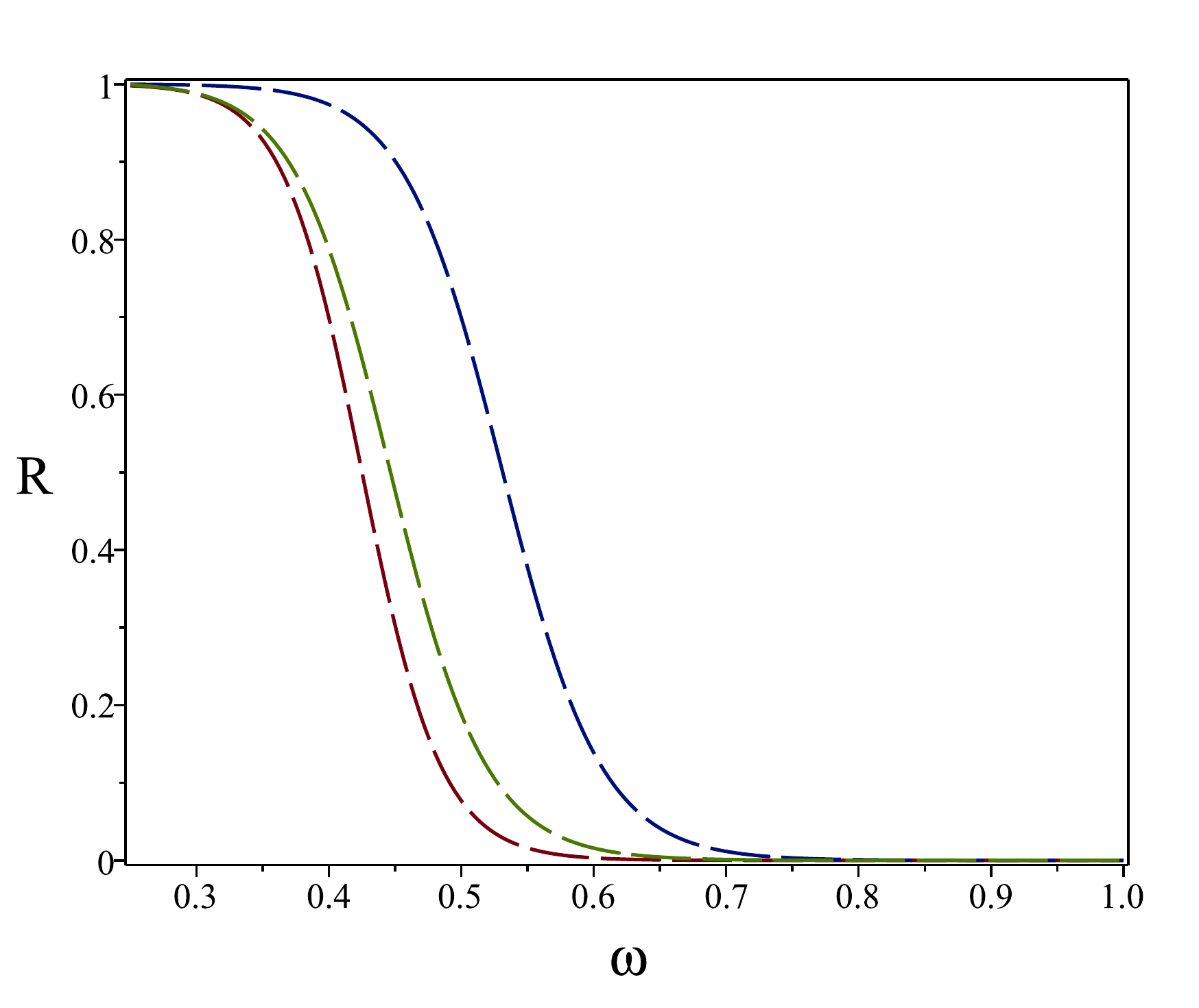}
\caption{\small \textit{The reflection coefficient against $\omega$. We have chosen $M=1$, $\xi=0.2$, $m=2$, for the red line; $M=1$, $\xi=0.25$, $m=2$ for the green line, and $M=1.5$, $\xi=0.25$, $m=2$, for the blue line. }}
\end{figure}

Note that the reflection coefficient goes to zero at the limit of large $\omega$ as shown in Fig.1 which means that a black hole absorbs more at high energies than the smaller energies.

\subsection{Flux calculation and greybody factors}
\begin{figure}[h!]
\center
\includegraphics[width=0.45\textwidth]{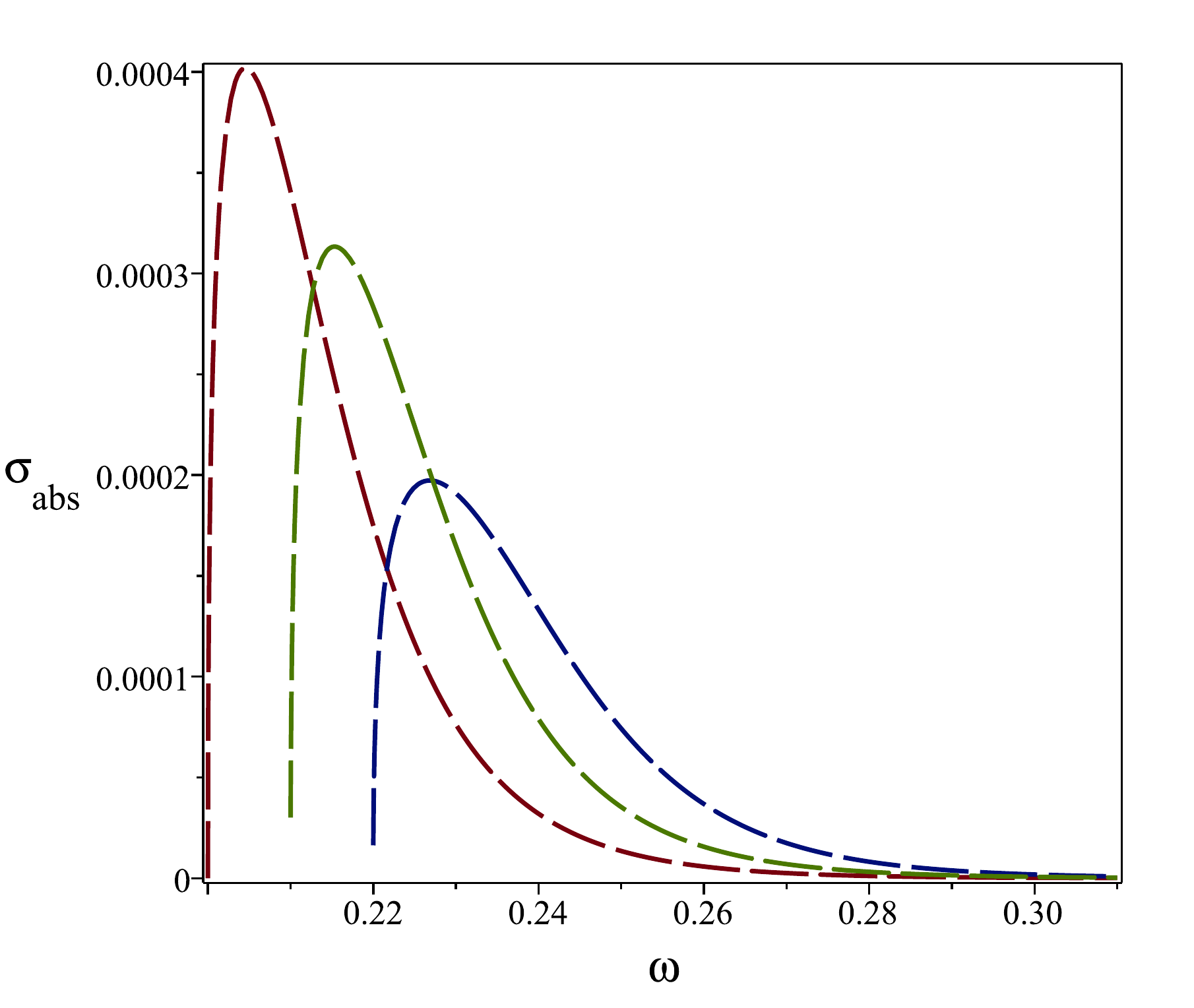}
\caption{\small \textit{The absorption cross section against $\omega$ is plotted for three cases. We have chosen $M=1$, $\xi=0.2$, $m=1.4$; $M=1.1$, for the red curve; $\xi=0.22$, $m=1.6$;$M=1.2$, for the green curve; and $\xi=0.21$, $m=1.58$, for the blue curve. }}
\end{figure}
Here, the flux of the ingoing/outgoing waves is studied to obtain the absorption cross section at the horizon and infinity. For this purpose, we define the conserved flux
\begin{equation}
\mathcal{F} = \frac{2 \pi} {i} \left ( R^{*} f(r) \frac{dR(r)}{dr} - R(r) f(r) \frac{ dR^*}{dr} \right),
\end{equation}
where $f(r)= 2\xi( r-r_{h})$. After we use the $R(r)$ at the horizon, the ingoing flux becomes,
\begin{equation}
\mathcal{F}(r \rightarrow r_h) = 4 \pi |A_1|^2 \omega.
\end{equation}

Then we use Eq. (\ref{rr}) to find the incoming flux at infinity as follows:
\begin{equation}
\mathcal{F}(r \rightarrow \infty) = 4 \pi  \omega |D_2|^2 \sqrt{1-\frac{\xi^2}{\omega^2}}.
\end{equation}
 The partial wave absorption (greybody factor) is defined as \cite{qnm98},
\begin{equation}
\sigma = \frac{ \mathcal{F}(r \rightarrow r_h)}{\mathcal{F}(r \rightarrow \infty)}.
\end{equation}

By substituting the above expression simplifies to,
\begin{equation}
\sigma=\frac{\omega}{\sqrt{\omega^2-\xi^2}}|\frac{\Gamma(a) \Gamma(b)}{\Gamma(c) \Gamma(a+b-c)}|^2.
\end{equation}

Using the values of parameters $a,b,c$ we can rewrite the above result also as
\begin{equation}
\sigma=\frac{\sinh\left(\frac{\pi \omega}{\xi}\right) \sinh\left(\frac{\pi \sqrt{\omega^2-\xi^2}}{\xi}\right)}{\cosh[\pi (\frac{\omega+\sqrt{\omega^2-\xi^2}}{\xi}+\frac{m}{\sqrt{M}})]\cosh[\pi (\frac{\omega+\sqrt{\omega^2-\xi^2}}{\xi}-\frac{m}{\sqrt{M}})]}.
\end{equation}

The greybody factor (the absorption cross section) in $2+1$ dimensions is also written as $ \sigma_{abs} = \sigma/\omega$ \cite{qnm75,Das:1996we}. Hence, the absorption cross section is obtained as follows:
\begin{eqnarray}\notag
\sigma_{abs} &=& \frac{\left(e^{\frac{2\pi \omega}{\xi}}-1\right)\left(e^{\frac{2\pi \sqrt{\omega^2-\xi^2}}{\xi}}-1\right)}{\omega \left(e^{2\pi (\frac{\omega+\sqrt{\omega^2-\xi^2}}{\xi}+\frac{m}{\sqrt{M}})}+1\right)}\\
&\times & \frac{1}{\left(e^{2\pi (\frac{\omega+\sqrt{\omega^2-\xi^2}}{\xi}-\frac{m}{\sqrt{M}})}+1\right)},
\end{eqnarray}
and it is plotted in Fig.2. 

\subsection{Decay rates}

Then the decay rate is obtained as follows:
\begin{equation}
\Gamma_{decay} = \frac{\sigma_{abs}} { e^{\omega/T} -1}.
\end{equation}

The black hole temperature is found to be 
\begin{equation} \label{hr}
T_{H} =\frac{\xi}{2\pi}.
\end{equation}

Therefore the decay rate for this black hole is found as follows:
\begin{eqnarray}\notag
\Gamma_{decay} &=&\frac{\left(e^{\frac{2\pi \sqrt{\omega^2-\xi^2}}{\xi}}-1\right)}{\omega \left(e^{2\pi (\frac{\omega+\sqrt{\omega^2-\xi^2}}{\xi}+\frac{m}{\sqrt{M}})}+1\right)}\\
&\times & \frac{1}{\left(e^{2\pi (\frac{\omega+\sqrt{\omega^2-\xi^2}}{\xi}-\frac{m}{\sqrt{M}})}+1 \right)},
\end{eqnarray}
 where we plot the the decay rate against $\omega$ in Fig.3.

\begin{figure}[h!]
\center
\includegraphics[width=0.49\textwidth]{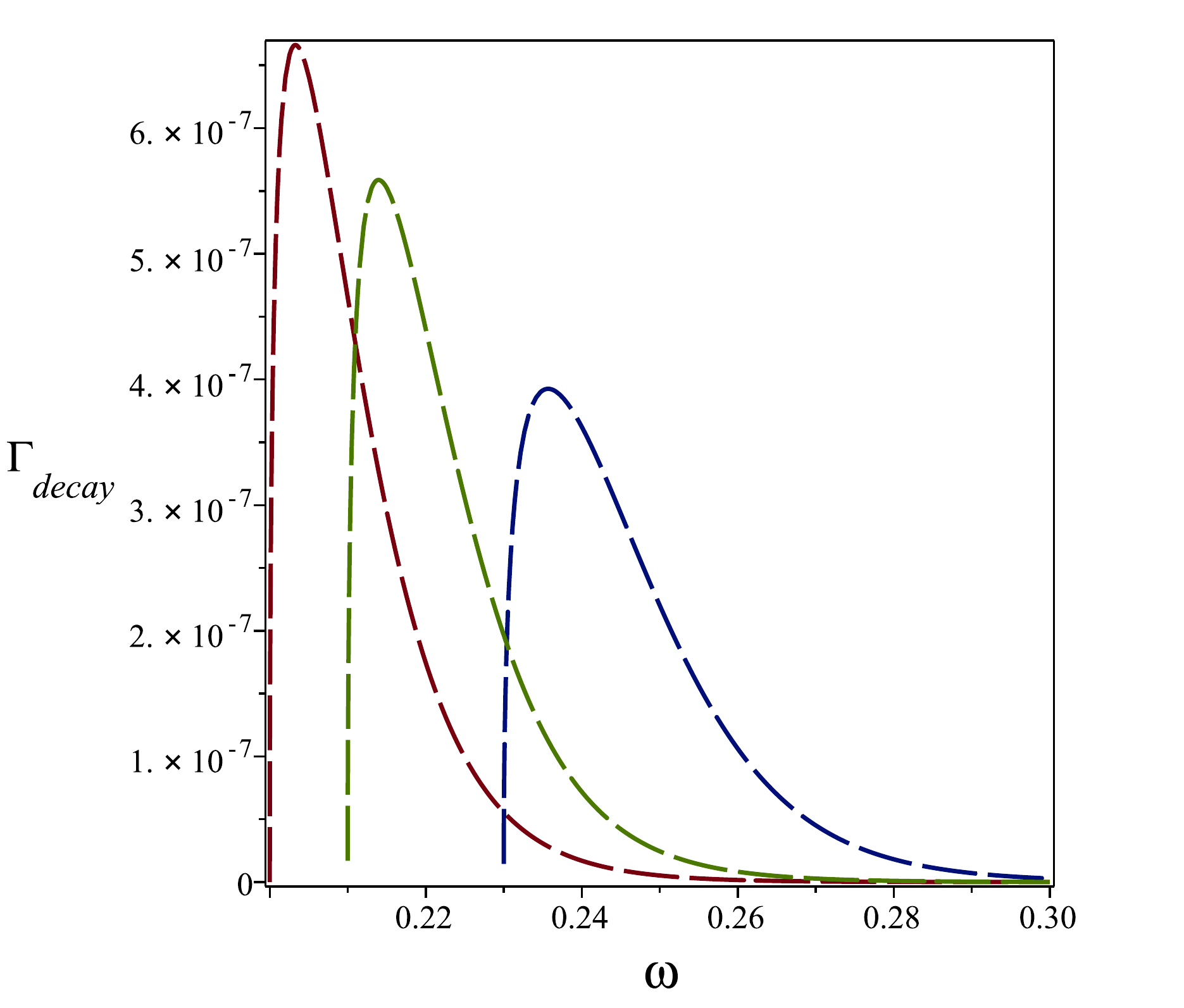}
\caption{\small \textit{We plot the decay rate against $\omega$ for three cases. We have chosen $M=1$, $\xi=0.2$, $m=1.4$, for the red curve; $M=1.1$, $\xi=0.22$, $m=1.6$, fpr the green curve; $M=1.2$, $\xi=0.21$, $m=1.58$, for the blue curve.}}
\end{figure}

\subsection{QNMs of a scalar field in the $2+1$ dimensional CSBHs}

To calculate the QNMs of the scalar field in the $2+1$ dimensional CSBHs, first we use the condition of Dirichlet, where the amplitude of the ingoing wave is taken zero ($D_1=0$). Then, using the vanishing of the incoming flux from infinity $ F_{\infty}$, we obtain $|D_1|^2=0$. Afterwards, the QNMs are calculated as follows:

\begin{equation}
D_1 = \frac{ \Gamma(c) \Gamma(c-a-b)}{\Gamma(c-a) \Gamma(c-b)}, 
\end{equation}
It is noted that the function $\Gamma(x)$ has poles at $c-a=-n$ and $c-b=-n$ for $ n=0,1,2..$. For the condition of vanishing $D_1$ at these poles, we find this relation:
\begin{equation}
 \frac{1}{2}- i\left( \frac{\omega+\sqrt{\omega^2-\xi^2}}{2 \xi} \mp \frac{m}{\sqrt{M}} \right) = -n.
\end{equation}

In order to find stable solution we impose the following condition
\begin{equation} \label{67}
\sqrt{1-\frac{\xi^2}{\omega^2}} >0,
\end{equation}
provided $\omega>\xi$. Having stable QNMs, one must have $\text{Im} \omega<0$, hence by combining two equations given by Eq. (\ref{67}) we find
\begin{eqnarray}
\omega_{n} &=& -2\,i\,\frac{ n \left(n+1\right)\xi}{2n+1}.
\end{eqnarray}

Having stable QNMs, one must have $\text{Im} \omega<0$, hence the negative sign in the last equation clearly indicates stability in the propagation of a scalar
field in background of $2+1$ dimensional CSBH. However choosing $\beta=\beta_{-}$, it is possible to obtain a new set of QNMs
\begin{equation}
 \frac{1}{2}+ i\left( \frac{\omega+\sqrt{\omega^2-\xi^2}}{2 \xi} \mp \frac{m}{\sqrt{M}} \right) = -n,
\end{equation}
which then lead to 
\begin{eqnarray} \label{omegaa}
\omega_{n} &=& 2\,i\,\frac{ n \left(n+1\right)\xi}{2n+1}.
\end{eqnarray}

Which means an instability of the propagation of a scalar field in this case. To this end, we shall accept only the first case, namely the stable solution as a physical solution. 

Let us use Eq. (\ref{omegaa}) to calculate the transition frequency as follows
\begin{equation}
\Delta\omega =\text{Im}\left(\omega_{n-1}-\omega_{n}\right)=\frac{4 n^2 \xi}{4n^2-1}.
\end{equation}

In particular for highly damped modes this relation becomes
\begin{equation}
\Delta\omega =\lim_{n\to \infty}\left(\frac{4 n^2 \xi}{4n^2-1}\right)=\xi,
\end{equation}
where the surface gravity is given by
\begin{equation}
\kappa=\frac{2 \pi T_{H}}{\hbar}.
\end{equation}

Note that in the last equation we have introduced the Planck constant. Furthermore from Eq. (\ref{hr}) we see that $\kappa=\xi$. Considering the adiabatic invariance  which is defined as
\begin{equation}
I_{adb}=\int\frac{T_{H}\Delta S_{BH}}{\Delta \omega},
\end{equation}
at this point, we make use of the quantization condition which basically says that $I_{adb,n}=n\hbar$, thus the entropy is found to be
\begin{equation}
S_{BH,n}=2\pi n.
\end{equation}

Finally, we use this result to find the area spectrum as follows
\begin{equation}
A_{BH,n}= 4\hbar S_{BH,n}=8\pi \hbar n.
\end{equation}

The last result is in perfect agreement with the Bekenstein's conjecture $[A_n=\epsilon n \hbar]$ \cite{Bekenstein1,Bekenstein2}. Yielding the minimum  change to the area of the horizon $\Delta A_{min}=A_n-A_{n-1}=8\pi \hbar$  with $\epsilon=8\pi$. 

\section{Conclusion}
In this paper we have calculated QNMs and greybody factors of a massless scalar field in the background of $2+1$ dimensional CSBH exactly. Firstly, we have performed exact analytic solution to the field equations and computed the reflection coefficient, absorption cross section, decay rate in terms of the hypergeometric functions, and also Hawking temperature. Secondly, we have analyzed QNMs and found a stable solution in the case when $\beta=\beta_{+}$ satisfying the equation \begin{eqnarray} \label{omegr}
\omega_{n} &=& -2\,i\,\frac{ n \left(n+1\right)\xi}{2n+1}.
\end{eqnarray}. 

Contrary to this case, we find an instability in the case when we choose $\beta=\beta_{-}$. The calculations of QNMs are calculated analytically and show the effect of the cloud of strings on the QNMs as follows: \begin{eqnarray}
\omega_{n} &=& 2\,i\,\frac{ n \left(n+1\right)\xi}{2n+1}.
\end{eqnarray}
Only the stable solution make sense, in this way we can say that the stable QNMs relation (\ref{omegr}) can be accepted as a physical solutions.

 \begin{acknowledgments}
This work was supported by Comisi\'on Nacional de Ciencias y Tecnolog\'ia of Chile through FONDECYT Grant N$^{\textup{o}}$ 3170035 (A. \"{O}.).  A. \"{O}. is grateful to Prof. Douglas Singleton for hosting him as a
research visitor at the California State University, Fresno. In addition A. 
\"{O}. would like to thank Prof. Leonard Susskind and Stanford Institute for
Theoretical Physics for hospitality.
\end{acknowledgments}


\begin{thebibliography}{99}

\bibitem{hawking1}
S. W. Hawking, Nature (London) 248, 30 (1974).

 \bibitem{hawking2}Commun. Math. Phys. 43, 199 (1975); 46, 206(E) (1976).
 
 \bibitem{Bekenstein1} 
  J.~D.~Bekenstein,
  Phys.\ Rev.\ D {\bf 7}, 2333 (1973).
  
  \bibitem{Bekenstein2} 
  J.~D.~Bekenstein,
  Phys.\ Rev.\ D {\bf 12}, 3077 (1975).
  
\bibitem{wil1} 
  M.~K.~Parikh and F.~Wilczek,
  Phys.\ Rev.\ Lett.\  {\bf 85}, 5042 (2000).
  
\bibitem{par1}M. K. Parikh, Gen.\ Rel.\ Grav. \textbf{36}, 2419 (2004).

\bibitem{corda1}
  C.~Corda,
  Int.\ J.\ Mod.\ Phys.\ D {\bf 21}, 1242023 (2012).
  
  \bibitem{corda2} 
  C.~Corda, S.~H.~Hendi, R.~Katebi and N.~O.~Schmidt,
  JHEP {\bf 1306}, 008 (2013).
  
  \bibitem{corda3} 
  C.~Corda, S.~H.~Hendi, R.~Katebi and N.~O.~Schmidt,
  Adv.\ High Energy Phys.\  {\bf 2014}, 527874 (2014).
  
  \bibitem{Kubiznak1} 
  D.~Kubiznak, R.~B.~Mann and M.~Teo,
  Class.\ Quant.\ Grav.\  {\bf 34}, no. 6, 063001 (2017).
  
  \bibitem{Kubiznak2} 
  D.~Kubiznak and R.~B.~Mann,
  Can.\ J.\ Phys.\  {\bf 93}, no. 9, 999 (2015).
  
 
  \bibitem{ovgun0} 
  I.~Sakalli, A.~\"{O}vg\"{u}n and K.~Jusufi,
  Astrophys.\ Space Sci.\  {\bf 361}, no. 10, 330 (2016).
  
  \bibitem{ovgun1} 
  A.~\"{O}vg\"{u}n and K.~Jusufi,
  Eur.\ Phys.\ J.\ Plus {\bf 131}, no. 5, 177 (2016).
  
  \bibitem{ovgun2} 
  A.~\"{O}vg\"{u}n and K.~Jusufi,
  Eur.\ Phys.\ J.\ Plus {\bf 132}, no. 7, 298 (2017).
  
  \bibitem{ovgun3} 
  K.~Jusufi, A.~Ovgun and G.~Apostolovska,
  Adv.\ High Energy Phys.\  {\bf 2017}, 8798657 (2017).
  
  \bibitem{ovgun4} 
  A.~\"{O}vg\"{u}n and K.~Jusufi,
  Adv.\ High Energy Phys.\  {\bf 2017}, 1215254 (2017).
  
  \bibitem{ovgun5} 
  K.~Jusufi and A.~\"{O}vg\"{u}n,
  Int.\ J.\ Theor.\ Phys.\  {\bf 56}, no. 6, 1725 (2017).
  
  
    \bibitem{mann1} 
  R.~Kerner and R.~B.~Mann,
  Phys.\ Rev.\ D {\bf 75}, 084022 (2007).
  
  \bibitem{mann2} 
  R.~Kerner and R.~B.~Mann,
  Phys.\ Lett.\ B {\bf 665}, 277 (2008).
  
  \bibitem{mann3} 
  R.~Kerner and R.~B.~Mann,
  Class.\ Quant.\ Grav.\  {\bf 25}, 095014 (2008).
  
  \bibitem{mann4} 
  R.~Kerner and R.~B.~Mann,
  Phys.\ Rev.\ D {\bf 73}, 104010 (2006).
  
  
  
  \bibitem{ovgun6} 
  X.~M.~Kuang, J.~Saavedra and A.~\"{O}vg\"{u}n,
  Eur.\ Phys.\ J.\ C {\bf 77}, no. 9, 613 (2017).
  
  \bibitem{sin1} 
  S.~K.~Modak and D.~Singleton,
  Eur.\ Phys.\ J.\ C {\bf 75}, no. 5, 200 (2015).
  
  \bibitem{sin2} 
  A.~Zampeli, D.~Singleton and E.~C.~Vagenas,
  JHEP {\bf 1206}, 097 (2012).
  
  \bibitem{sin3} 
  S.~K.~Modak and D.~Singleton,
  Int.\ J.\ Mod.\ Phys.\ D {\bf 21}, 1242020 (2012).
  
  \bibitem{sin4} 
  D.~Singleton and S.~Wilburn,
  Phys.\ Rev.\ Lett.\  {\bf 107}, 081102 (2011).
  
  \bibitem{sin5} 
  T.~Zhu, J.~R.~Ren and D.~Singleton,
  Int.\ J.\ Mod.\ Phys.\ D {\bf 19}, 159 (2010).
  
  \bibitem{sin6}
  E.~T.~Akhmedov, V.~Akhmedova and D.~Singleton,
  Phys.\ Lett.\ B {\bf 642}, 124 (2006).
  
  \bibitem{sin7} 
  E.~T.~Akhmedov, V.~Akhmedova, T.~Pilling and D.~Singleton,
  Int.\ J.\ Mod.\ Phys.\ A {\bf 22}, 1705 (2007).
  

  
    
  \bibitem{bir1} 
  T.~Birkandan and M.~Hortacsu,
  Gen.\ Rel.\ Grav.\  {\bf 35}, 457 (2003).
  
  \bibitem{bir2} 
  T.~Birkandan and M.~Hortacsu,
  J.\ Math.\ Phys.\  {\bf 48}, 092301 (2007).
  
  \bibitem{bir3} 
  T.~Birkandan and M.~Hortacsu,
  J.\ Phys.\ A {\bf 40}, 1105 (2007).
  
  \bibitem{bir4} 
  T.~Birkandan and M.~Cvetic,
  JHEP {\bf 1409}, 121 (2014).
  
  \bibitem{bir5} 
  T.~Birkandan and M.~Hortacsu,
  EPL {\bf 119}, no. 2, 20002 (2017).
  
  \bibitem{bir6} 
  T.~Birkandan and M.~Hortacsu,
  arXiv:1711.06811 [gr-qc].
  
     \bibitem{sak1} 
  I.~Sakalli and A.~\"{O}vg\"{u}n,
  Eur.\ Phys.\ J.\ Plus {\bf 131}, no. 6, 184 (2016).
  
  
   
  
    \bibitem{Hod1} 
  S.~Hod,
  Phys.\ Rev.\ Lett.\  {\bf 81}, 4293 (1998).
  
  \bibitem{Hod2} 
  S.~Hod,
  Phys.\ Rev.\ D {\bf 67}, 081501 (2003).
    
  \bibitem{Hod3} 
  S.~Hod,
  Class.\ Quant.\ Grav.\  {\bf 23}, L23 (2006).
  
    
  
  \bibitem{Hod4} 
  U.~Keshet and S.~Hod,
  Phys.\ Rev.\ D {\bf 76}, 061501 (2007).
  
  
  
  \bibitem{Hod5} 
  S.~Hod,
  Phys.\ Rev.\ D {\bf 75}, 064013 (2007).
 
   \bibitem{Hod6} 
  S.~Hod,
  Phys.\ Lett.\ B {\bf 666}, 483 (2008).
  
  
    
  \bibitem{Hod7} 
  S.~Hod,
  Phys.\ Rev.\ D {\bf 78}, 084035 (2008).
  
    
  \bibitem{Hod8} 
  S.~Hod,
  Phys.\ Rev.\ D {\bf 80}, 064004 (2009).
  
  \bibitem{Hod9} 
  S.~Hod,
  Phys.\ Rev.\ D {\bf 84}, 044046 (2011).
  
  
   
  \bibitem{Hod10} 
  S.~Hod,
  Phys.\ Lett.\ B {\bf 761}, 53 (2016).
  
 \bibitem{Mag} 
  M.~Maggiore,
  Phys.\ Rev.\ Lett.\  {\bf 100}, 141301 (2008).
  
  \bibitem{qnm0} A. L\'{o}pez-Ortega, Int. J. Mod. Phys. D \textbf{18},
1441-1459 (2009).


\bibitem{qnm1} 
  A.~\"{O}vg\"{u}n, I.~Sakalli and J.~Saavedra,
  arXiv:1708.08331.
  
\bibitem{qnm2} 
  K.~Jusufi, I.~Sakallı and A.~\"{O}vg\"{u}n,
  Gen.\ Rel.\ Grav.\  {\bf 50}, no. 1, 10 (2018).
  
\bibitem{qnm3} 
  P.~A.~Gonzalez, A.~\"{O}vg\"{u}n, J.~Saavedra and Y.~Vasquez,
  arXiv:1711.01865 [gr-qc].
  

\bibitem{qnm4} 
  P.~A.~Gonzalez, E.~Papantonopoulos, J.~Saavedra and Y.~Vasquez,
  Phys.\ Rev.\ D {\bf 95}, no. 6, 064046 (2017).
  
\bibitem{qnm5} 
  M.~Cruz, M.~Gonzalez-Espinoza, J.~Saavedra and D.~Vargas-Arancibia,
  Eur.\ Phys.\ J.\ C {\bf 76}, no. 2, 75 (2016).
  
\bibitem{qnm6} 
  P.~A.~Gonzalez, J.~Saavedra and Y.~Vasquez,
  Int.\ J.\ Mod.\ Phys.\ D {\bf 21}, 1250054 (2012).
  
\bibitem{qnm7} 
  R.~Becar, S.~Lepe and J.~Saavedra,
  Int.\ J.\ Mod.\ Phys.\ A {\bf 25}, 1713 (2010).
  
\bibitem{qnm8} 
  P.~Gonzalez, E.~Papantonopoulos and J.~Saavedra,
  JHEP {\bf 1008}, 050 (2010).
  
\bibitem{qnm9} 
  R.~Becar, S.~Lepe and J.~Saavedra,
  J.\ Phys.\ Conf.\ Ser.\  {\bf 134}, 012007 (2008).
  
\bibitem{qnm10} 
  R.~Becar, S.~Lepe and J.~Saavedra,
  Phys.\ Rev.\ D {\bf 75}, 084021 (2007).
  
\bibitem{qnm11} 
  J.~Saavedra,
  Mod.\ Phys.\ Lett.\ A {\bf 21}, 1601 (2006).
  
 
  

  
\bibitem{qnm12} 
  S.~Lepe and J.~Saavedra,
  Phys.\ Lett.\ B {\bf 617}, 174 (2005).
  
\bibitem{qnm13} 
  J.~Crisostomo, S.~Lepe and J.~Saavedra,
  Class.\ Quant.\ Grav.\  {\bf 21}, 2801 (2004).
  
\bibitem{qnm14} 
  R.~Becar, P.~A.~Gonzalez, J.~Saavedra and Y.~Vasquez,
  Eur.\ Phys.\ J.\ C {\bf 75}, no. 2, 57 (2015).
  
    \bibitem{qnm15} 
  G.~Kunstatter,
  Phys.\ Rev.\ Lett.\  {\bf 90}, 161301 (2003).


\bibitem{qnm16} 
  C.~Campuzano, P.~Gonzalez, E.~Rojas and J.~Saavedra,
  JHEP {\bf 1006}, 103 (2010).
  
\bibitem{qnm17} 
  P.~Gonzalez, E.~Papantonopoulos and J.~Saavedra,
  JHEP {\bf 1008}, 050 (2010).
  
\bibitem{qnm18} 
  R.~Becar, P.~A.~Gonzalez and Y.~Vasquez,
  Eur.\ Phys.\ J.\ C {\bf 74}, no. 8, 3028 (2014).
  
\bibitem{qnm19} 
  R.~Aros, F.~Bugini and D.~E.~Diaz,
    J.\ Phys.\ Conf.\ Ser.\  {\bf 720}, no. 1, 012009 (2016).

    
\bibitem{qnm20} 
  R.~Aros, C.~Martinez, R.~Troncoso and J.~Zanelli,
  Phys.\ Rev.\ D {\bf 67}, 044014 (2003).
  
\bibitem{qnm21} 
  B.~Wang, E.~Abdalla and R.~B.~Mann,
  Phys.\ Rev.\ D {\bf 65}, 084006 (2002).
  
\bibitem{qnm22} 
  I.~Sakalli,
  Int.\ J.\ Mod.\ Phys.\ A {\bf 26}, 2263 (2011)
  Erratum: [Int.\ J.\ Mod.\ Phys.\ A {\bf 28}, 1392002 (2013)].
  
\bibitem{qnm23} 
  I.~Sakalli,
  Mod.\ Phys.\ Lett.\ A {\bf 28}, 1350109 (2013).
  
\bibitem{qnm24} 
  I.~Sakalli and S.~F.~Mirekhtiary,
  Astrophys.\ Space Sci.\  {\bf 350}, 727 (2014).
  
\bibitem{qnm25} 
  I.~Sakalli,
  Eur.\ Phys.\ J.\ C {\bf 75}, no. 4, 144 (2015).
  
\bibitem{qnm26} 
  I.~Sakalli and G.~Tokgoz,
  Adv.\ High Energy Phys.\  {\bf 2015}, 739153 (2015).
  
\bibitem{qnm27} 
  I.~Sakalli and G.~Tokgoz,
  Annalen Phys.\  {\bf 528}, 612 (2016).
  
  
\bibitem{qnm28} 
  I.~Sakalli,
  Phys.\ Rev.\ D {\bf 94}, no. 8, 084040 (2016).
  
\bibitem{qnm29} 
  I.~Sakalli and O.~A.~Aslan,
  Astropart.\ Phys.\  {\bf 74}, 73 (2016).
  
\bibitem{qnm30} 
  G. Panotopoulos and A. Rincon,
 doi:10.1142/S0218271818500347
  arXiv:1711.04146 [hep-th].
  
  \bibitem{qnm31} 
  A.~F.~Cardona and C.~Molina,
  Class.\ Quant.\ Grav.\  {\bf 34}, no. 24, 245002 (2017).
  
  \bibitem{qnm32} 
  M.~Wang, C.~Herdeiro and J.~Jing,
  Phys.\ Rev.\ D {\bf 96}, no. 10, 104035 (2017).
  
  \bibitem{qnm33} 
  A.~Jansen,
  Eur.\ Phys.\ J.\ Plus {\bf 132}, no. 12, 546 (2017).
  
  \bibitem{qnm34} 
  J.~Zhang and S.~Y.~Zhou,
  arXiv:1709.07503 [gr-qc].
  
  \bibitem{qnm35} 
  R.~Dong, J.~Sakstein and D.~Stojkovic,
  Phys.\ Rev.\ D {\bf 96}, no. 6, 064048 (2017).
  
  \bibitem{qnm36} 
  D.~Q.~Sun, Z.~L.~Wang, M.~He, X.~R.~Hu and J.~B.~Deng,
  Adv.\ High Energy Phys.\  {\bf 2017}, 4817948 (2017).
  
  \bibitem{qnm37} 
  C.~Chirenti,
  doi:10.1007/s13538-017-0543-7.
  
  \bibitem{qnm38} 
  B.~Toshmatov and Z.~Stuchlík,
  Eur.\ Phys.\ J.\ Plus {\bf 132}, no. 7, 324 (2017).
  
  \bibitem{qnm39} 
  C.~Ding,
  Phys.\ Rev.\ D {\bf 96}, no. 10, 104021 (2017).
  
  \bibitem{qnm40} 
  K.~D.~Kokkotas, R.~A.~Konoplya and A.~Zhidenko,
  Phys.\ Rev.\ D {\bf 96}, no. 6, 064004 (2017).
  
  \bibitem{qnm41} 
  J.~L.~Blazquez-Salcedo, F.~S.~Khoo and J.~Kunz,
  Phys.\ Rev.\ D {\bf 96}, no. 6, 064008 (2017).
  
  \bibitem{qnm42} 
  R.~A.~Konoplya and A.~Zhidenko,
  JHEP {\bf 1709}, 139 (2017).
  
  \bibitem{qnm43} 
  R.~A.~Konoplya and Z.~Stuchlík,
  Phys.\ Lett.\ B {\bf 771}, 597 (2017).
  
  \bibitem{qnm44} 
  B.~Toshmatov, C.~Bambi, B.~Ahmedov, Z.~Stuchlík and J.~Schee,
  Phys.\ Rev.\ D {\bf 96}, 064028 (2017).
  
  \bibitem{qnm45} 
  S.~H.~Volkel and K.~D.~Kokkotas,
  Class.\ Quant.\ Grav.\  {\bf 34}, no. 17, 175015 (2017).
  
  \bibitem{qnm46} 
  S.~Bhattacharyya and S.~Shankaranarayanan,
  Phys.\ Rev.\ D {\bf 96}, no. 6, 064044 (2017).
  
  \bibitem{qnm47} 
  J.~Matyjasek and M.~Opala,
  Phys.\ Rev.\ D {\bf 96}, no. 2, 024011 (2017).
  
  
  \bibitem{qnm48} 
  E.~Berti, V.~Cardoso and A.~O.~Starinets,
  Class.\ Quant.\ Grav.\  {\bf 26}, 163001 (2009).
  

  
    \bibitem{qnm49} 
  D.~Birmingham, S.~Carlip and Y.~j.~Chen,
  Class.\ Quant.\ Grav.\  {\bf 20}, L239 (2003).
  
    \bibitem{qnm50} 
  R.~A.~Konoplya and A.~Zhidenko,
  Rev.\ Mod.\ Phys.\  {\bf 83}, 793 (2011).
  
  \bibitem{qnm51} 
  O.~Dreyer,
  Phys.\ Rev.\ Lett.\  {\bf 90}, 081301 (2003).
  
  \bibitem{qnm52} 
  R.~A.~Konoplya,
  Phys.\ Rev.\ D {\bf 68}, 024018 (2003).
  
  \bibitem{qnm53} 
  V.~Cardoso and J.~P.~S.~Lemos,
  Phys.\ Rev.\ D {\bf 63}, 124015 (2001).
  
  \bibitem{qnm54} 
  V.~Cardoso and J.~P.~S.~Lemos,
  Phys.\ Rev.\ D {\bf 64}, 084017 (2001).
  
  \bibitem{qnm55} 
  V.~Cardoso, R.~Konoplya and J.~P.~S.~Lemos,
  Phys.\ Rev.\ D {\bf 68}, 044024 (2003).
  
  \bibitem{qnm56} 
  V.~Ferrari and B.~Mashhoon,
  Phys.\ Rev.\ D {\bf 30}, 295 (1984).
  
  \bibitem{qnm57} 
  T.~Andrade, J.~Casalderrey-Solana and A.~Ficnar,
  JHEP {\bf 1702}, 016 (2017).
  
  \bibitem{qnm58} 
  T.~Andrade, A.~Castro and D.~Cohen-Maldonado,
  Class.\ Quant.\ Grav.\  {\bf 34}, no. 9, 095009 (2017).
  
  \bibitem{qnm59} 
  P.~K.~Kovtun and A.~O.~Starinets,
  Phys.\ Rev.\ D {\bf 72}, 086009 (2005).
  
  \bibitem{qnm60} 
  N.~Andersson,
  Phys.\ Rev.\ D {\bf 55}, 468 (1997).
  
  \bibitem{qnm61} 
  B.~Chen and Z.~b.~Xu,
  JHEP {\bf 0911}, 091 (2009).
  
  \bibitem{qnm62} 
  B.~Wang, C.~Y.~Lin and E.~Abdalla,
  Phys.\ Lett.\ B {\bf 481}, 79 (2000).
  
  \bibitem{qnm63} 
  B.~Wang, C.~Y.~Lin and C.~Molina,
  Phys.\ Rev.\ D {\bf 70}, 064025 (2004).
  
  \bibitem{qnm64} 
  D.~P.~Du, B.~Wang and R.~K.~Su,
  Phys.\ Rev.\ D {\bf 70}, 064024 (2004).

  
  \bibitem{qnm65} 
  J.~M.~Maldacena and A.~Strominger,
  Phys.\ Rev.\ D {\bf 55}, 861 (1997).
  
  \bibitem{qnm66} 
  P.~Kanti and J.~March-Russell,
  Phys.\ Rev.\ D {\bf 67}, 104019 (2003).
  
  \bibitem{qnm67} 
  M.~Cvetic and F.~Larsen,
  Nucl.\ Phys.\ B {\bf 506}, 107 (1997).
  
  \bibitem{qnm68} 
  M.~Cvetic and F.~Larsen,
  JHEP {\bf 0909}, 088 (2009).
  
  \bibitem{qnm69} 
  J.~Ahmed and K.~Saifullah,
  Eur.\ Phys.\ J.\ C {\bf 77}, no. 12, 885 (2017).
  
  \bibitem{qnm70} 
  C.~Y.~Zhang, P.~C.~Li and B.~Chen,
  arXiv:1712.00620 [hep-th].
  
  \bibitem{qnm71} 
  G.~Panotopoulos and A.~Rincon,
  Phys.\ Rev.\ D {\bf 96}, no. 2, 025009 (2017).

\bibitem{qnm72} 
  G.~Panotopoulos and A.~Rincon,
  Phys.\ Lett.\ B {\bf 772}, 523 (2017).
  
  \bibitem{qnm73} 
  R.~Dong and D.~Stojkovic,
  Phys.\ Rev.\ D {\bf 92}, no. 8, 084045 (2015).
  
  \bibitem{qnm74} 
  X.~M.~Kuang and J.~P.~Wu,
  Phys.\ Lett.\ B {\bf 770}, 117 (2017).


  
  \bibitem{qnm75} 
  S.~Fernando,
  Gen.\ Rel.\ Grav.\  {\bf 37}, 461 (2005).
  
  \bibitem{qnm76} 
  S.~Fernando,
  Gen.\ Rel.\ Grav.\  {\bf 36}, 71 (2004).
  
  \bibitem{qnm77} 
  S.~Fernando and K.~Arnold,
  Gen.\ Rel.\ Grav.\  {\bf 36}, 1805 (2004).
  
  \bibitem{qnm78} 
  S.~Fernando,
  Gen.\ Rel.\ Grav.\  {\bf 37}, 585 (2005).
  
  \bibitem{qnm79} 
  S.~Fernando,
  Int.\ J.\ Mod.\ Phys.\ A {\bf 25}, 669 (2010).
  
  \bibitem{qnm80} 
  S.~Fernando,
  Phys.\ Rev.\ D {\bf 77}, 124005 (2008).
  
  \bibitem{qnm81} 
  S.~Fernando,
  Phys.\ Rev.\ D {\bf 79}, 124026 (2009).
  
  \bibitem{qnm82} 
  S.~Fernando and J.~Correa,
  Phys.\ Rev.\ D {\bf 86}, 064039 (2012).
  
  \bibitem{qnm83} 
  S.~Fernando and T.~Clark,
  Gen.\ Rel.\ Grav.\  {\bf 46}, no. 12, 1834 (2014).
  
  \bibitem{qnm84} 
  S.~Fernando,
  Mod.\ Phys.\ Lett.\ A {\bf 30}, no. 11, 1550057 (2015).
  
  \bibitem{qnm85} 
  S.~Fernando,
  Mod.\ Phys.\ Lett.\ A {\bf 30}, no. 13, 1550078 (2015).
  
  \bibitem{qnm86} 
  S.~Fernando,
  Mod.\ Phys.\ Lett.\ A {\bf 30}, no. 29, 1550147 (2015).
  
  \bibitem{qnm87} 
  S.~Fernando,
  Int.\ J.\ Mod.\ Phys.\ D {\bf 24}, no. 14, 1550104 (2015).
  
  \bibitem{qnm88} 
  S.~Fernando,
  Gen.\ Rel.\ Grav.\  {\bf 48}, no. 3, 24 (2016).
  
  \bibitem{qnm89} 
  S.~Fernando and A.~Manning,
  Int.\ J.\ Mod.\ Phys.\ D {\bf 26}, no. 09, 1750100 (2017).
  
  \bibitem{qnm90} 
  N.~Breton, T.~Clark and S.~Fernando,
  Int.\ J.\ Mod.\ Phys.\ D {\bf 26}, no. 10, 1750112 (2017).
  
  \bibitem{qnm91} 
  J.~P.~Morais Graca, G.~I.~Salako and V.~B.~Bezerra,
  Int.\ J.\ Mod.\ Phys.\ D {\bf 26}, no. 10, 1750113 (2017).
  
  \bibitem{qnm92} 
  J.~P.~Morais Graca and I.~P.~Lobo,
  arXiv:1711.08714 [gr-qc].
  
  \bibitem{qnm93} 
  J.~P.~Morais Graca, H.~S.~Vieira and V.~B.~Bezerra,
  Gen.\ Rel.\ Grav.\  {\bf 48}, no. 4, 38 (2016).
  

 


  
  \bibitem{qnm94} 
  M.~R.~Setare and E.~C.~Vagenas,
  Mod.\ Phys.\ Lett.\ A {\bf 20}, 1923 (2005).
  
  \bibitem{qnm95} 
  E.~C.~Vagenas,
  JHEP {\bf 0811}, 073 (2008).
  
 
  \bibitem{qnm96} 
  D.~Birmingham, I.~Sachs and S.~N.~Solodukhin,
  Phys.\ Rev.\ Lett.\  {\bf 88}, 151301 (2002).
  
  \bibitem{qnm97} P. Kanti, T. Pappas, N. Pappas, Phys. Rev. D 90, 124077.
  
  \bibitem{qnm98} T. Harmark, J. Natario, R. Schiappa, 	Adv. Theor. Math. Phys. 14 (2010) 727-793.
  
  \bibitem{Schiappa1} V. Cardoso, J. Natario, R. Schiappa, J.Math.Phys. 45 (2004) 4698-4713.
  
  \bibitem{Schiappa2} J. Natario, R. Schiappa, Adv.Theor.Math.Phys.8:1001-1131,2004.
  
  \bibitem{liu1} D. Zou, Y. Liu, C. Zhang, B. Wang, Europhys.Lett. 116 (2016) no.4, 40005.
  
 \bibitem{liu2} D. Zou, Y. Liu, R. Yue, Eur. Phys. J. C (2017) 77:365.
 
 \bibitem{liu3} Y. Liu, D. Zou, B. Wang, J. High Energ. Phys. (2014) 2014: 179.
  
 
    \bibitem{LIGO} 
  B.~P.~Abbott {\it et al.} [LIGO Scientific and Virgo Collaborations],
  Phys.\ Rev.\ Lett.\  {\bf 116}, no. 22, 221101 (2016).
  
  
  
  
  
  \bibitem{Banados:1992wn} 
  M.~Banados, C.~Teitelboim and J.~Zanelli,
  Phys.\ Rev.\ Lett.\  {\bf 69}, 1849 (1992).
  
\bibitem{Mazharimousavi:2015sfo} 
  S.~H.~Mazharimousavi and M.~Halilsoy,
  Eur.\ Phys.\ J.\ C {\bf 76}, no. 2, 95 (2016).
  
 \bibitem{ricardo} J. Oliva, D. Tempo, R. Troncoso, JHEP 0907:011, 2009. 
  
\bibitem{Das:1996we} 
  S.~R.~Das, G.~W.~Gibbons and S.~D.~Mathur,
  Phys.\ Rev.\ Lett.\  {\bf 78}, 417 (1997).
  
\bibitem{Horowitz:1993jc} 
  G.~T.~Horowitz and D.~L.~Welch,
  Phys.\ Rev.\ Lett.\  {\bf 71}, 328 (1993).
  
  
  \bibitem{Chandrasekhar:1975zza} 
  S.~Chandrasekhar and S.~L.~Detweiler,
  Proc.\ Roy.\ Soc.\ Lond.\ A {\bf 344}, 441 (1975).
  
  \bibitem{zib} 
  D.~B.~Sibandze, R.~Goswami, S.~D.~Maharaj, A.~M.~Nzioki and P.~K.~S.~Dunsby,
  Eur.\ Phys.\ J.\ C {\bf 77}, no. 6, 364 (2017).
  
  
  

\end{thebibliography}
\end{document}